\documentclass[aps,prl,twocolumn,amsmath,amssymb,floatfix,longbibliography]{revtex4-2}
\usepackage{amsmath,amssymb}
\usepackage{bm}
\usepackage{braket}   
\usepackage{graphicx}
\usepackage{xcolor}
\usepackage{cancel}   
\usepackage[colorlinks=true,allcolors=blue]{hyperref}

\begin{document}

\title{Coherence in Biological Systems}

\author{Yehuda Roth}
\affiliation{Science Department, Oranim College of Education, Tivon 3600600, Israel}

\begin{abstract}
	When does a collection of autonomous cells become a multicellular individual?
	We propose that coherence provides a physical description of this transition.
	Coherence is treated as a global property arising when distinguishable
	constituents admit a physically meaningful collective state-space description.
	Using the center of mass and an interaction-based construction, we show that
	such collective states can be defined for classical bodies before dynamics is
	introduced, with normal modes emerging as a particular dynamical realization.
	We apply this framework to multicellular organization, where cells retain
	their identities while their independent individuality is replaced by
	participation in the organized whole. In \emph{Dictyostelium discoideum},
	cAMP-mediated coupling produces population-level collective modes, while
	starvation provides an experimentally controlled energetic constraint on the
	transition to multicellularity. The framework yields direct tests through
	interaction-derived collective eigenstates and the energetic cost of
	maintaining autonomous versus collective organization. Coherence may thus
	provide a general physical description of multicellular individuality without
	requiring microscopic quantum coherence or intrinsic wave character.
\end{abstract}

\maketitle

\section{Introduction}

A central problem in the emergence of multicellularity is the
\emph{transition in individuality}: how do autonomous cells become
constituents of a higher-level individual? Evolutionary accounts describe
this transition through cooperation, division of labor, communication,
interdependence, and the suppression of within-group conflict
\cite{MichodRoze2001,West2015}. Here we address the same transition from a
physical perspective.

Coherence is inherently a global property. Whether in classical waves or
quantum systems, it characterizes relations among the constituents that
permit a description of the system as a whole; it is not a property assigned
independently to each constituent. This makes coherence a natural concept for
multicellular organization. A multicellular organism consists of physically
distinct cells that retain their identities, yet they no longer function as
independent biological individuals. Their physical and biological roles are
defined within an organized whole. The transition from autonomous cells to
such constituents therefore suggests a corresponding physical transition:
from a constituent-level description to a collective state-space description
of the system.

We refer to this property as \emph{system-level coherence}. Unlike classical
wave coherence, including optical coherence of electromagnetic fields
\cite{MandelWolf1995}, or microscopic quantum coherence, the proposed
coherence applies to systems composed of distinguishable classical bodies.
Its defining feature is not synchronization or collective motion alone, but
the existence of a physically meaningful collective state associated with
the whole. The constituents retain their identities while their independent
individuality is replaced by participation in the collective description.

The purpose of this work is to develop the physical and mathematical basis
for this proposal. Our central ordering is \emph{coherence before dynamics}:
collective states are first selected by relations among the constituents,
while dynamics subsequently determines how those states are realized or
maintained. The center of mass provides a first example; an interaction
operator provides a second, with normal-mode dynamics emerging only after a
spatial realization is introduced. We then apply the resulting collective
state-space construction to multicellular organization, using the starvation
response of \emph{Dictyostelium discoideum} as an existing biological
example.

\section{Classical coherence: The center-of-mass coordinate}
The next two sections establish the classical basis of the proposed
construction. In both, the basis $\{|k\rangle\}$ labels classical bodies and
coherence is represented by a linear combination of constituent states. The
center of mass shows how a collective observable fixes the participation
coefficients. We then show that an interaction operator selects collective
eigenstates before any equation of motion is specified. Dynamics enters only
after a spatial realization is introduced. Thus linearity belongs to the
collective representation, not necessarily to the underlying dynamics.

Classical mechanics provides a transparent example of a physically meaningful
quantity belonging to a system as a whole. In the state-space representation
used here, the state vector is not itself regarded as a directly measurable
quantity. Its physical content is expressed through expectation values of
operators representing measurable quantities. We therefore construct the
state associated with a classical system by requiring that these expectation
values reproduce the corresponding classical observables.

Consider a system $S$ composed of $N$ particles, where $|k\rangle$ labels
particle $k$ and $m_k$ denotes its mass. We begin with the normalized state
\begin{equation}
	|I_S\rangle
	=
	\sum_{k\in S}A_k|k\rangle,
	\qquad
	\sum_{k\in S}|A_k|^2=1 .
	\label{eq:impact_general}
\end{equation}

Let $q$ denote a spatial degree of freedom and define
\begin{equation}
	\mathbb Q_S
	=
	\sum_{k\in S}q_k|k\rangle\langle k| .
	\label{eq:qoperator}
\end{equation}
Its expectation value is
\begin{equation}
	\langle I_S|\mathbb Q_S|I_S\rangle
	=
	\sum_{k\in S}|A_k|^2q_k .
	\label{eq:qexpectation}
\end{equation}
To recover the classical center-of-mass coordinate,
\begin{equation}
	q_{\rm CM}
	=
	\frac{1}{M_S}\sum_{k\in S}m_kq_k,
	\qquad
	M_S=\sum_{k\in S}m_k,
	\label{eq:classical_com}
\end{equation}
comparison with Eq.~(\ref{eq:qexpectation}) requires
\begin{equation}
	|A_k|^2=\frac{m_k}{M_S}.
	\label{eq:relative_impact}
\end{equation}
Thus the center-of-mass construction provides a first example of system-level
coherence defined independently of dynamics: a collective state is specified
by the constituent states, while the classical observable fixes their
mass-dependent participation coefficients.

\section{Interaction-selected collective states}

Consider $N$ interacting classical bodies represented by the constituent
basis ${|k\rangle}$. Before introducing positions, time, or equations of
motion, we define the dimensionless interaction operator in the constituent
representation as
\begin{equation}
	  \mathbb {K}_{\{k\}}
	=
	\sum_{k,j=1}^{N}
	K_{kj}|k\rangle\langle j| .
	\label{eq:interaction_constituent}
\end{equation}
At this stage, the dimensionless matrix elements $K_{kj}$ characterize the
interaction structure among the constituents without specifying the resulting
dynamics or introducing a physical force scale.

The collective states selected by the interaction are defined by
\begin{equation}
	  \mathbb {K}_{\{k\}}|\alpha\rangle
	=
	\lambda_\alpha|\alpha\rangle .
	\label{eq:interaction_eigenvalue}
\end{equation}
Writing
\begin{equation}
	|\alpha\rangle
	=
	\sum_{k=1}^{N}
	A_k^{(\alpha)}|k\rangle,
	\qquad
	\langle\alpha|\alpha\rangle=1 ,
	\label{eq:interaction_collective_state}
\end{equation}
the coefficients $A_k^{(\alpha)}$ specify the relative participation of the
constituents in the collective eigenstate.

The important point is that
Eqs.~(\ref{eq:interaction_constituent})--
(\ref{eq:interaction_collective_state}) contain no position, time, velocity,
acceleration, or force scale. The interaction has therefore selected
collective states before the dynamics is introduced. According to the present
proposal, this collective representation defines the coherence of the
interacting bodies. When dynamics is subsequently introduced, these states
may acquire a dynamical manifestation as normal modes.

We now perform the realization of the same interaction operator in the
spatial representation. Let $\mathbb U_q$ denote the unitary transformation
from the constituent representation to the spatial representation,

The interaction operator transforms according to
\begin{equation}
	  \mathbb {K}_{\{k\}}
	\longrightarrow
	 \mathbb {K}_{\{q\}},
	\qquad
	 \mathbb {K}_{\{q\}}
	=
	\mathbb U_q
	  \mathbb {K}_{\{k\}}
	\mathbb U_q^\dagger .
	\label{eq:interaction_spatial}
\end{equation}

Since the transformation is unitary, the eigenvalue structure is preserved.
Defining
\begin{equation}
	|\alpha\rangle_{{q}}
	=
	\mathbb U_q|\alpha\rangle ,
	\label{eq:collective_spatial}
\end{equation}
we obtain
\begin{equation}
	 \mathbb {K}_{\{q\}}|\alpha\rangle_{{q}}
	=
	\lambda_\alpha|\alpha\rangle_{{q}} .
	\label{eq:interaction_spatial_eigenvalue}
\end{equation}

The realization therefore does not generate the collective state. It
expresses in the spatial representation a collective state that was already
selected by the interaction in the constituent representation.

In general, the individual spatial states $|q_k\rangle$ need not be
eigenstates of $ \mathbb {K}_{\{q\}}$. The interaction may therefore couple the
spatial states, whereas its eigenstates describe collective combinations of
the interacting bodies.

Only now do we introduce dynamics. Since $\mathbb K_{\{q\}}$ is
dimensionless, a physical scale is required for its Newtonian realization.
We therefore define
\begin{equation}
	\mathbb F^{\rm int}_{\{q\}}
	=
	-F_0\mathbb K_{\{q\}} ,
	\label{eq:operator_mechanics}
\end{equation}
where $F_0$ is a characteristic force scale of the particular physical
system. It is not a universal constant and does not enter the definition of
coherence. Rather, $\mathbb K_{\{q\}}$ determines the dimensionless
interaction structure, while $F_0$ supplies its physical strength.

In an oscillatory linear realization, $F_0$, together with the characteristic
mass and length scales of the system, determines the frequency scale of the
resulting normal modes. The collective eigenstates and their dimensionless
eigenvalues are therefore defined before the dynamical time scale is
introduced.

In general, the interaction-force operator may depend on the spatial
coordinates,
\begin{equation}
	\mathbb F^{\rm int}_{\{q\}}
	=
	\mathbb F^{\rm int}_{\{q\}}(q_1,\ldots,q_N),
\end{equation}
and is represented by a non-diagonal matrix with elements
\begin{equation}
	a_{k,k'}
	=
	\langle q_k|
	\mathbb F^{\rm int}_{\{q\}}
	|q_{k'}\rangle .
\end{equation}
The off-diagonal elements describe the coupling between different constituents, whereas the diagonal elements determine the force acting on each constituent,
\begin{equation}
	m_k\ddot q_k
	=
	\langle q_k|
	\mathbb F^{\rm int}_{\{q\}}
	|q_k\rangle .
	\label{eq:individual_force}
\end{equation}
For small deviations about a given configuration, the interaction-force
operator may be linearized. In this approximation, the collective
eigenstates acquire the standard dynamical realization as classical
normal modes.

This establishes the logical ordering central to the present construction:
\begin{equation}
	\begin{aligned}
	\text{interaction}
	&\longrightarrow
	\text{collective state}\\
	&\longrightarrow
	\text{spatial realization}
	\longrightarrow
	\text{dynamics}.
\end{aligned}
	\label{eq:coherence_ordering}
\end{equation}

\section{Identity and collective organization in multicellular systems}
The classical constructions above separate constituent identity from
participation in the collective whole. We apply the same distinction to cells.

Let $\nu$ denote the host genetic code and associate cell $k$ with
\begin{equation}
 |\nu\rangle_k .
 \label{eq:hostcode}
\end{equation}
Genetic identity labels the constituent \cite{Roth2026ClassicalAging}; it
does not by itself establish coherence or membership. A colony of autonomous
cells is therefore represented component-wise,
\begin{equation}
 |N\rangle_{\rm col}
 =
 \bigotimes_{k=1}^{N}|\nu\rangle_k.
 \label{eq:colony}
\end{equation}
Guided by our central assumption, and supported structurally by the classical examples above, we propose that system-level coherence in a multicellular organism is represented by a collective superposition of its constituent states,
\begin{equation}
 |N\rangle_{\rm org}
 =
 \sum_{k=1}^{N}A_k|\nu\rangle_k ,
 \qquad
 \sum_{k=1}^{N}|A_k|^2=1 ,
 \label{eq:organism}
\end{equation}
where $A_k$ characterizes relative participation in the collective
description, without a probabilistic interpretation.

Here, superposition refers to a linear state-space representation of classical bodies and does not imply microscopic quantum superposition. 

The distinction is organizational, not genetic: the same identity coordinate
appears in both representations. Autonomous cells are product factors;
organismal cells participate in the collective state. Thus identity specifies
what the constituent is, membership whether it belongs to the chosen system,
and $A_k$ how it participates. 

The construction thus gives physical content to the biological distinction
between a collection and an organism: the former is constituent-defined,
whereas the latter admits a collective system-level representation.
In this sense, system-level coherence supplies a state-space description of
the transition in individuality: a cell changes from an autonomous biological
unit to a constituent whose physical role is defined within the organized
whole.

\section{Biological realization in \emph{Dictyostelium}}

The social amoeba \emph{Dictyostelium discoideum} provides a biological system
in which to examine the present construction. Under nutrient-rich conditions,
cells behave as autonomous amoebae, whereas starvation induces multicellular
development \cite{Gregor2010,Kelly2021}. Starving cells communicate through
extracellular cyclic adenosine monophosphate (cAMP), producing synchronized
population-level oscillations and propagating waves that coordinate
aggregation \cite{Gregor2010}. Diffusive cAMP coupling stabilizes these
oscillations against cell-to-cell variability \cite{Kim2007}, and propagating
cAMP waves persist in multicellular mounds and slugs \cite{Singer2019}.

Within the present framework, cAMP is not coherence itself but a physical
coupling between constituent cells, analogous to the coupling between
classical oscillators. The population-level oscillations are then a dynamical
realization of the collective structure. A direct test would be to construct
an effective interaction matrix from measured cAMP-mediated coupling,
determine its collective eigenstates, and compare them with the observed
population-level modes. Importantly, organized collective migration persists
after cAMP oscillations and their propagation disappear \cite{Hashimura2019},
supporting the distinction between collective organization and a particular
dynamical realization.

Starvation also introduces an energetic constraint. Kelly \emph{et al.}
showed that nutrient limitation reduces mitochondrial respiration and
proliferation through a sulfur-dependent metabolic switch; inhibiting
mitochondrial respiration accelerates aggregation, whereas essential amino
acids restore mitochondrial activity and suppress starvation-induced
aggregation \cite{Kelly2021}. This suggests that collective organization may
be favored under restricted resources if its maintenance requires less power
than maintaining the same constituents as autonomous units,
\begin{equation}
	P_{\rm collective}<\sum_{k=1}^{N}P_k .
	\label{eq:collective_power}
\end{equation}
This inequality is not a result of Ref.~\cite{Kelly2021}, but a testable
prediction of the present framework. Comparative measurements of the
metabolic power required to maintain autonomous and collective organization
under controlled conditions could test this selection criterion.

Thus \emph{D. discoideum} provides two experimental routes to the proposal:
comparison of interaction-derived eigenstates with collective modes, and a
direct energetic test of collective versus autonomous organization.

\section{Discussion and outlook}

The motivation for this work was the conceptual similarity between cells in
a multicellular organism and constituents of a coherent physical system: in
both cases, distinguishable constituents participate in an organization whose
physically relevant description belongs to the whole.

The classical constructions developed here provide a physical basis for this
correspondence. The center of mass demonstrates a collective state defined
independently of dynamics, while the interaction construction shows that
collective eigenstates can be selected before their dynamical realization.
Normal modes then provide a familiar mechanical manifestation of this
ordering.

Applied to multicellularity, coherence describes the transition from
autonomous cellular individuals to constituents of an organized whole,
without requiring the cells to lose their identities. The proposal is
experimentally accessible: collective eigenstates inferred from
cAMP-mediated coupling can be compared with observed population-level modes,
while the energetic selection criterion can be tested by comparing the power
required to maintain autonomous and collective organization.

Coherence therefore need not be restricted to quantum systems or classical
waves. It can characterize classical systems of distinguishable bodies when
their physically relevant description becomes collective. Multicellular
organization provides a natural biological realization of this general
physical concept.

\begin{acknowledgments}
The author thanks Professor Yoram Gershman for helpful advice and discussions.
OpenAI ChatGPT (GPT-5.6 Sol) was used for literature exploration,
mathematical consistency checks, manuscript organization, and language
revision. The author independently evaluated the output, verified the
scientific claims and references, and assumes full responsibility for the
manuscript.
\end{acknowledgments}

\bibliography{sources_PRL}

\end{document}